\documentclass[epj,twocolumn,floatfix,final]{svjour}
\usepackage{graphicx}

\begin{document}
\title{Finite-well potential in the 3D  nonlinear
Schr\"odinger
equation: Application to
Bose-Einstein condensation  }

\author{Sadhan K. Adhikari\thanks{e-mail: adhikari@ift.unesp.br}}
\institute{Instituto de F\'{\i}sica Te\'orica, UNESP $-$ S\~ao Paulo 
State
University,  01.405-900 S\~ao Paulo, S\~ao Paulo, Brazil}

\date{Received: date / Revised version: date}
%

\abstract{Using variational and numerical solutions we show that 
stationary
negative-energy localized (normalizable) bound states can appear in the
three-dimensional nonlinear Schr\"odinger equation with a finite
square-well potential for a range of nonlinearity parameters. Below a
critical attractive nonlinearity, the system becomes unstable and
experiences collapse.  Above a limiting repulsive nonlinearity, the
system becomes highly repulsive and cannot be bound.  The system also
allows nonnormalizable states of infinite norm at positive energies in
the continuum.  The normalizable negative-energy bound states could be
created in BECs and studied in the laboratory with present knowhow.}

\PACS{{45.05.+x}{General theory of classical mechanics of discrete
systems} \and {05.45.-a}{Nonlinear dynamics and chaos} \and
{03.75.Hh}{Static properties of condensates; thermodynamical,
statistical, and structural properties}}

 \authorrunning{S. K. Adhikari}
\titlerunning{Finite-well potential in the 3D  nonlinear
Schr\"odinger
equation}
 
%
\maketitle

\section{Introduction}

The nonlinear Schr\"odinger (NLS) equation with cubic or Kerr 
nonlinearity
appears in many areas of
physics and mathematics \cite{1}. Of these,
two areas have drawn much
attention in
recent time. They are pulse propagation in nonlinear medium
\cite{0,0a,3DNLS} and
Bose-Einstein condensation (BEC) in confining traps \cite{8}. A
quantum-mechanical
mean-field description of BEC is done using the nonlinear
Gross-Pitaevskii
(GP) equation
which is essentially the NLS equation with cubic nonlinearity.
Though the NLS equation in
these two areas have similar mathematical structure, the interpretation
of the different terms in it is quite distinct. In BEC it is an 
extension
of
the Schr\"odinger equation to include a nonlinear interaction among
bosons. In optics it describes electromagnetic pulse propagation
in a nonlinear medium.  Also, usually there is no
external potential in the NLS equation in optics \cite{1}, whereas in
BEC a trapping potential is to be included in it \cite{8}. In most 
studies
in BEC an infinite parabolic harmonic
potential has been  included in
the NLS equation which simulates the infinite or nearly infinite
experimental parabolic
magnetic trap.

In this paper we consider a finite square-well trapping
potential in the NLS equation with cubic nonlinearity. Although, we
consider a square-well potential for obvious analytical knowledge about
this potential, most of our results should be valid for any finite
potential and the experiments are really carried out on finite traps.
 This potential is
piecewise constant and
leads to analytic solution in
many one-dimensional (1D) problems, and serves as a model for a trap of
finite
depth and can
be
realized in the laboratory. Because of these features,
there have been several studies of the 1D
NLS equation with a finite
square-well and other simple potentials. Zakharov and Shabat \cite{0a}
found the
solutions to the NLS equation with a constant potential on the infinite
line. Carr {\it et al.} solved the NLS equation under periodic and box
boundary conditions \cite{c1} as well as with the 1D
square-well
potential \cite{c2}. There have been studies  of the step function
\cite{step,step2},
point-like impurity potential \cite{pt} and the
parabolic potential \cite{para}
in the NLS
equation, as well as transmission of matter wave across various
potentials  \cite{tr}.

We extend the above 1D
investigations to the spherically-symmetric
three-dimensional (3D) square-well interaction. This is possibly the
most-studied problem in linear quantum mechanics and allows analytic
or quasi-analytic
treatment in many cases. Also, it models an
experimental situation
which can be realized
with present-day BEC technology with the use of a detuned laser beam of
finite intensity \cite{c2,step2}. Such an optical device  could generate 
a
square-well
potential in
one
direction \cite{c2}. Three such  potentials in orthogonal directions
could make an excellent model for a finite 3D square-well
potential. In BEC three standing-wave orthogonal laser beams have 
already
been used to make a 3D periodic optical-lattice potential
\cite{ol}.
 In view of this, the creation of a 3D square-well potential
seems possible.
Once a BEC is materialized in a square-well potential,
it could be studied in the laboratory and the results
compared with the prediction of the
theoretical models based on the GP
equation, thus
providing    stringent tests for these
models.

We show that it is possible to have normalizable
stationary  BEC bound states in
localized finite 3D square-well potentials
for a range of nonlinearity parameters. A too strongly attractive
nonlinearity
parameter is found to lead to collapse, whereas a very strong nonlinear
repulsion does not bind the system.  In addition to the
normalizable stationary bound states, the repulsive NLS equation with
square-well
interaction is also found to yield  nonnormalizable  states
where
the probability density has a central peaking on a constant background
extending to infinity. Obviously, these  nonnormalizable
states
do not satisfy the boundary condition that the wave function $\psi(r)$ 
at
a radial distance $r$ should tend to 0 as $r \to \infty$.
The formation and the study of the  normalizable states
could be of utmost interest
in several areas, e.g., optics
\cite{1}, nonlinear physics \cite{1} and BEC \cite{8}, whereas the
nonnormalizable states  will draw the attention of
researchers in mathematical and nonlinear physics. We
use both variational as well as numerical solutions of the
NLS equation in our study.

In this connection we mention that in an exponentially decaying finite
potential well one could have the interesting possibility of quantum
tunneling and the appearance of quasi-bound states, which has been
studied in detail in Ref. \cite{malo}. In the present study with
square-well
potential this possibility is not of
concern.

In Sec. II we present the theoretical
model which we use in our
investigation.
In Sec. III we explain how to obtain numerically the usual normalizable
solutions of the NLS equation with the finite and infinite square-well
potentials. We also explain the origin of the nonnormalizable solutions
and how to obtain them numerically. Then
we develop a time-dependent variational method for the study of this
problem. The nonlinear
problem is reduced to an
effective potential well.
The
possibility of the appearance of stable
bound states in this effective potential for a wide range of the
parameters is discussed.
In Sec. IV  we consider the
complete numerical solution of the NLS equation for a finite and 
infinite
square-well potentials.
We obtain the condition of stability
of these bound states numerically and find their wave functions.  We
also obtain the nonnormalizable solutions of the NLS equation 
numerically.
Finally,  in Sec. V  a brief summary is given.

\section{The Nonlinear Schr\"odinger Equation}

We begin with  the radially-symmetric time-dependent
qu\-antum-mechanical GP equation  used to describe a BEC at 0 K
\cite{8}.
As we shall not be concerned with a particular
experimental system, we write the GP equation in dimensionless
variables.
The radial part of the
3D spherically-symmetric GP equation for the
Bose-Einstein condensate wave
function $\Phi({\cal R};\tau)$
at position ${\cal R}$ and time $\tau $ can be
written   as \cite{9}
\begin{eqnarray}
\biggr[- i\hbar \frac{\partial
}{\partial \tau} &-& \frac{\hbar^2}{2m} \biggr({\frac{\partial
^2}{\partial
{\cal R}^2} + \frac{2
}{\cal R}\frac{\partial }{\partial R}   } \biggr)
+ v({\cal R}) \nonumber \\
&+&
  G\left|
{\Phi({\cal  R};\tau)}\right|^2
 \biggr]\Phi({\cal  R};\tau)=  0,
\end{eqnarray}
where $G=4\pi \hbar^2 a N/m$ is the  nonlinearity, and $v({\cal R})$ is
the
square-well potential. Here $m$ is the mass of each atom, $N$ the number
of atoms  and $a$ is the atomic
scattering length. We introduce convenient dimensionless variables  by
$r={\cal R}/l$, $t=\tau\omega/2$, $\Psi=\Phi l^{3/2}$,
$V(r)=2v({\cal R})/(\hbar\omega)$, where
$l=\sqrt{\hbar/(m\omega)}$ and where  $\omega$
is an external reference angular frequency. In terms of these new
variables the GP equation becomes
\begin{eqnarray}\label{d1}
\biggr[- i\frac{\partial
}{\partial t} - {\frac{\partial
^2}{\partial
r^2} - \frac{2
}{r}\frac{\partial }{\partial r}   }
+ V(r) +
  g\left|
{\Psi({ r};t)}\right|^2\biggr]\Psi({  r};t)=  0,
\end{eqnarray}
where $g=8\pi a N/l$.
The square well potential is taken as
\begin{eqnarray} \label{po}
V(r)= -\gamma^2,\hskip .2cm  r \le \Lambda; \hskip 0.5cm
    = 0, \hskip .2cm  r>\Lambda,
\end{eqnarray}
with $\gamma^2$ the depth and $\Lambda$ the range.
Equation (\ref{d1}) with cubic nonlinearity
is the usual
NLS equation often used in problems of optics and nonlinear physics
and
will be referred to as the NLS equation in the following. If we set 
$g=0$ in
Eq. (\ref{d1}), this equation
becomes the usual linear Schr\"odinger equation.
In BEC $t$ denotes time and $r$ the space variable. In nonlinear
optics, $t$ denotes the direction of propagation of pulse, $r$ denotes
the transverse directions, and $\Psi$ refers to components of
electromagnetic field.  In nonlinear optics the 3D NLS equation is
spatiotemporal in nature where  for anomalos dispersion the time
variable can be  combined with the two
space variables in transverse directions to define the 3D vector $\vec 
r$
with $r=|\vec r|$. There have been many numerical studies of the 3D NLS
equation in optical pulse propagation \cite{3DNLS,9a}.
In BEC a  scaled nonlinearity $n$
is often defined by $n = g/(8\pi)=  { N}a/l.$
The normalization condition in Eq.   (\ref{d1}) is
\begin{equation} \label{n}
 \int d^3{ r} |\Psi({r};t)|^2 = 1.
\end{equation}

\section{Analytic  Consideration}

\subsection{Normalizable Solution}

The localized normalizable solutions to nonlinear equation
(\ref{d1}) with potential
(\ref{po})
are allowed only at
negative energies.
We  solve numerically Eq.  (\ref{d1}) starting from
a time iteration of the linear problem obtained by setting $g=0$ in this
equation. Hence we present a brief summary of the linear problem in the
following \cite{qm}. The stationary solution of the nonlinear equation
(\ref{d1}),
which we look for, has the form $\Psi({ r};t)= \psi(r) \exp(-i\mu t)$ 
with
$\mu$ the chemical potential, so that
\begin{eqnarray}\label{d2}
\biggr[-{\frac{\partial ^2}{\partial r^2} - \frac{2
}{r}\frac{\partial }{\partial r}   }
+ V(r)  +
  g\left|
{\psi({ r})}\right|^2
 \biggr]\psi({  r})=  \mu \psi(r).
\end{eqnarray}

For the piecewise constant square-well potential, the
solution of
the linear problem with $g=0$
is expressed in terms of the variable $\alpha =
\sqrt{(\gamma^2-|\mu|)}$, for $r\le \Lambda$; and
$\beta =\sqrt {|\mu|}$,
for
$r> \Lambda.$
In terms of these variables the solution of
Eq. (\ref{d2}) regular at the origin is expressed as
\begin{eqnarray}\label{w1}
\psi(r)&=& A \frac{\sin(\alpha r)}{\alpha r},\hskip .3cm r\le \Lambda, 
\\
&=& -\frac{B}{\beta r}\exp(-\beta r),\hskip .3cm r> \Lambda. \label{w2}
\end{eqnarray}
These solutions are discrete and
normalizable satisfying Eq. (\ref{n})
corresponding to  a negative  $\mu$.
The unknown  parameter
$\mu$ is obtained by matching the reduced
wave function  $r\psi(r)$ and its derivative at
$r=\Lambda$:
\begin{equation}\label{bc}
\alpha \Lambda \cot(  \alpha \Lambda)= -\beta  \Lambda.
\end{equation}
Once $\mu$ is known the wave function
$\psi(r)$ can also  be determined by implementing the normalization
condition (\ref{n}) of the wave function on Eqs. (\ref{w1}) and
(\ref{w2}).

The boundary condition (\ref{bc}) is simplified
for an infinite square-well potential, where $\sin(\alpha
\Lambda) =0$, so that $\alpha = j\pi/\Lambda$, where the integer $j=1$
corresponds to the ground state, and $j>1$   to the excited soliton
states. The
solution in this case is given by Eq. (\ref{w1}) with $\psi(r)=0$ for 
$r>
\Lambda
.$
After obtaining the solution of the linear equation with $g=0$ in this
fashion, the
nonlinear equation (\ref{d1}) is solved by time iteration.

\subsection{Nonnormalizable Solution}

The nonnormalizable solutions to Eq. (\ref{d2}) with the square-well
potential (\ref{po}) are only allowed for a
repulsive system for positive
$\mu$ values.  The stationary
bound-state solution of the linear problem discussed above  is
normalizable.
The solution of the nonlinear equation generated from that solution by
time iteration  is also normalizable. However, for positive $g$
(repulsive system)
Eq. (\ref{d2})
also has
nonnormalizable solutions with no counterpart in the linear
problem. They
are obtained
from time iteration of a special nonnormalizable solution of
Eq. (\ref{d2})
for $V(r)=0$. The normalization integral (\ref{n}) is now infinite
even for a system with finite number of particles  $N$ with
a finite scaled nonlinearity
$n$.
In other words a system with a finite $n$ (finite
scattering length $a$ and finite number of atoms $N$) can possess a wave
function with nonzero
probability density everywhere in space.

We note that  $\psi(r)=c$, a constant, is a
solution of Eq. (\ref{d2}) with $V(r)=0$, provided that
$\mu=gc^2$.
It is realized that for a repulsive system $\mu$
is positive and the
present state is a state in the continuum.
The required solution of Eq. (\ref{d2}) for a nonzero $V(r)$ is obtained
from this solution by time iteration while in each step of time 
iteration
the strength $\gamma^2$ of the square-well potential is increased slowly
until the desired value of $\gamma^2$ is attained and the final wave
function is obtained.  The final wave function tends towards the
nonzero constant $c$
for $r \to \infty$.

For a large condensate the kinetic energy term in Eq. (\ref{d1}) is
negligible and one has the following Thomas-Fermi (TF)  approximation to
the
wave function
\begin{equation}\label{tf1}
\psi_{\mbox{\small{TF}}}(r)=\frac{1}{\sqrt g} \sqrt{\mu-V(r)}.
\end{equation}
As $V(r)$ is piecewise constant, in the TF approximation
$\psi_{\mbox{\small{TF}}}(r)$ will also be piecewise constant. Usually, 
to
implement the TF
approximation we need the  parameter $|\mu|$. But in this
 case we
determine $\mu$ by requiring that in the TF approximation
$\psi_{\mbox{\small{TF}}}(r) \to c$
asymptotically. The actual numerical solution also tends to this
asymptotic limit.

If the solution of Eq. (\ref{d2}) is generated from the initial constant
solution $\psi(r)=c$, the TF approximation to the wave function becomes
\begin{eqnarray}\label{tf}
\psi_{\mbox{\small{TF}}}(r)\equiv C &=& \sqrt{(\mu+\gamma^2)/g}, \hskip
0.3cm r\le
\Lambda, \\
 \psi_{\mbox{\small{TF}}}(r) = c               &=&
\sqrt{\mu/g}, \hskip
0.3cm r>
\Lambda,
\end{eqnarray}
with
\begin{equation}
C=\sqrt{\frac{\gamma^2+c^2 g}{g}}.
\end{equation}

\subsection{Variational Result}

To understand how the normalizable bound states are formed in the
square-well potential  we employ a variational method with the following
Gaussian
wave function for the solution of Eq.  (\ref{d1})
\cite{9a}
\begin{equation}\label{twf}
\psi(r,t)=  A(t)\exp\left[-\frac{r^2}{2R^2(t)}
+\frac{i}{2}{ \beta(t) }r^2+i\alpha(t)
\right],
\end{equation}
where $A(t)\equiv [\pi^{3/4}R^{3/2}(t)]^{-1}$,  $R(t)$, $\beta(t)$, and
$\alpha(t)$ are the
normalization, width, chirp, and
phase of $\psi$, respectively. The Lagrangian density for
generating Eq.  (\ref{d1})   is \cite{9a,abdul}
\begin{eqnarray} {\cal L}(\psi)&=&\frac{i}{2}(\dot \psi\psi^* - \dot
\psi^* \psi )- \left|\frac{\partial
 \psi}{\partial r} \right|^2\nonumber \\ &  -&V(r)
|\psi|^2
-\frac{1}{2} g | \psi|^4 ,
\end{eqnarray}
where the overhead dot denotes time derivative.
The trial wave function (\ref{twf}) is
substituted in the Lagrangian density and the
effective Lagrangian $L_{\mbox{eff}}$
calculated by
integration: $L_{\mbox{eff}}= \int {\cal
L}(\psi)
d ^3 r.$
The Euler-Lagrange equations for this effective lagrangian are
 \cite{gold}
 \begin{equation}\label{el}
\frac{d}{d t}\frac{\partial L_{\mbox{eff}}}{\partial\dot q(t)}=
\frac{\partial L_{\mbox{eff}}}{\partial q(t)},
\end{equation}
where $q(t)$ stands for the generalized displacements   $R(t)$,
$\beta(t)$, $A(t)$ or $\alpha(t)$.

The following expression for the effective Lagrangian can be calculated
after some
straightforward algebra \cite{abdul}
\begin{eqnarray}
&L_{\mbox{eff}}&=\frac{\pi^{3/2}A^2(t)R^3(t)}{2}
\biggr[
-\frac{3}{2}\dot
\beta(t)
R^2(t)-\frac{1}{2\sqrt 2} g A^2(t)\nonumber \\
&-&2\dot \alpha(t)
-\frac{3}{R^2(t)}-3\beta^2(t) R^2(t)-
 {\cal F}_{\Lambda\gamma}(R)
\biggr],
\end{eqnarray}
where
\begin{equation}
{\cal F}_{\Lambda\gamma}(R)=\frac{8\gamma^2}{\sqrt \pi}
\left[
\frac{\Lambda}{2R}e^{-
\Lambda^2/R^2}         -\frac{\sqrt \pi}{4} \mbox{Erf}
\left( \frac{\Lambda}{R}
\right)
   \right],
\end{equation}
with the error function  Erf$(x)$ defined by
\begin{equation}
\mbox{Erf}(x)=\frac{2}{\sqrt \pi}\int_0^x e^{-t^2}dt.
\end{equation}
The Euler-Lagrange equations (\ref{el}) for $\alpha(t)$, $A(t), 
\beta(t),$
and
$R(t) $
are  given, respectively,  by
\begin{eqnarray}\label{e1}
\pi^{3/2}A^2R^3 =  \mbox{constant} &=& 1,  \\
\label{e2}
3\dot \beta + \frac{4\dot
\alpha}{R^2}+\frac{6}{R^4}
+\frac{2}{R^2}{\cal F}_{\Lambda\gamma}(R) + 6\beta^2
&=& -
\frac{2gA^2}{\sqrt
2R^2}, \\ \label{e3}
\dot R &=&  2 R \beta, \\
\label{e4}
5\dot \beta + \frac{4\dot
\alpha}{R^2}+\frac{2}{R^4}
+\frac{2}{R^2}{\cal F}_{\Lambda\gamma}(R)  &+& 10\beta^2
 \nonumber \\
+\frac{16}{3\sqrt
\pi}\frac{\gamma^2\Lambda^3}{R^5}e^{-\Lambda^2/R^2}
&=& -
\frac{gA^2}{\sqrt
2R^2},
\end{eqnarray}
where the time dependence of  different observables is  suppressed.
Eliminating $\alpha$ between  (\ref{e2}) and (\ref{e4}) one obtains
\begin{eqnarray}\label{e5}
2\dot \beta= \frac{4}{R^4}-4\beta^2+ \frac{gA^2}{\sqrt
2R^2}- \frac{16}{3\sqrt
\pi}\frac{\gamma^2\Lambda^3}{R^5}e^{-\Lambda^2/R^2}  .
\end{eqnarray}
From  (\ref{e3}) and  (\ref{e5}) we get the following second-order
differential equation for the evolution of the width $R$
\begin{eqnarray}\frac {d^2R }{dt^2}
&\equiv &
-\frac{dU(R)}{dR}\nonumber \\
&=&\frac{4}{R^3}+\frac{8n}{\sqrt{2
\pi}}\frac{1}{R^4}- \frac{16}{3\sqrt
\pi}\frac{\gamma^2\Lambda^3}{R^4}e^{-\Lambda^2/R^2} ,
\end{eqnarray}
where $n=g/(8\pi) $ and
$U(R)$ is the effective potential of motion given by
\begin{eqnarray}\label{ef1}
U(R)=
\frac{2}{R^2}+ \frac{8}{3\sqrt {2\pi}}\frac{n}{R^3}
+ \frac{2}{3}{\cal F}_{\Lambda\gamma}(R).
\end{eqnarray}
Small oscillation around a
stable configuration is possible when there is a minimum in this 
effective
potential denoting a stationary normalizable state.

\begin{figure}

\begin{center}
\includegraphics[width=.85\linewidth]{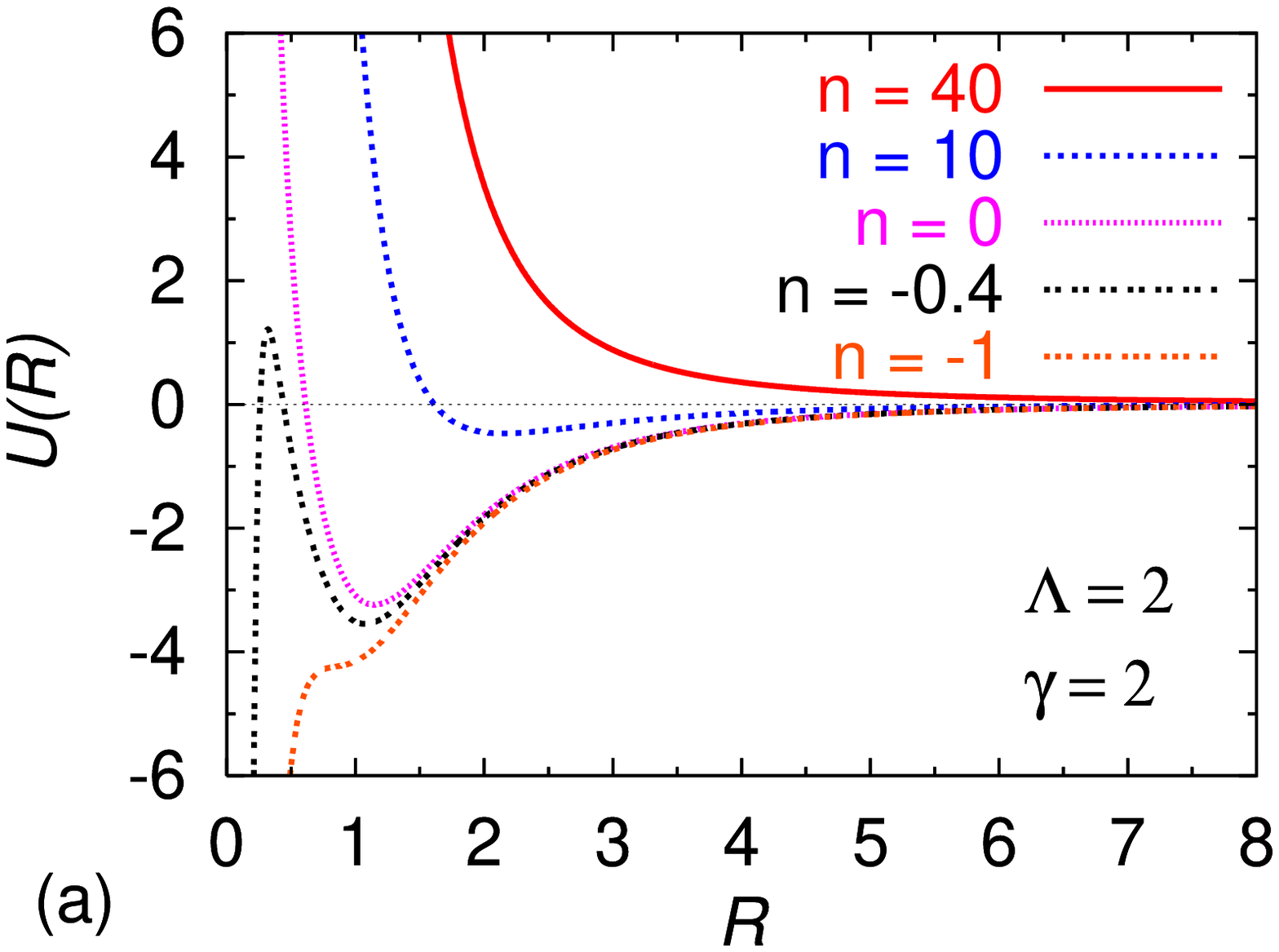}
\includegraphics[width=.85\linewidth]{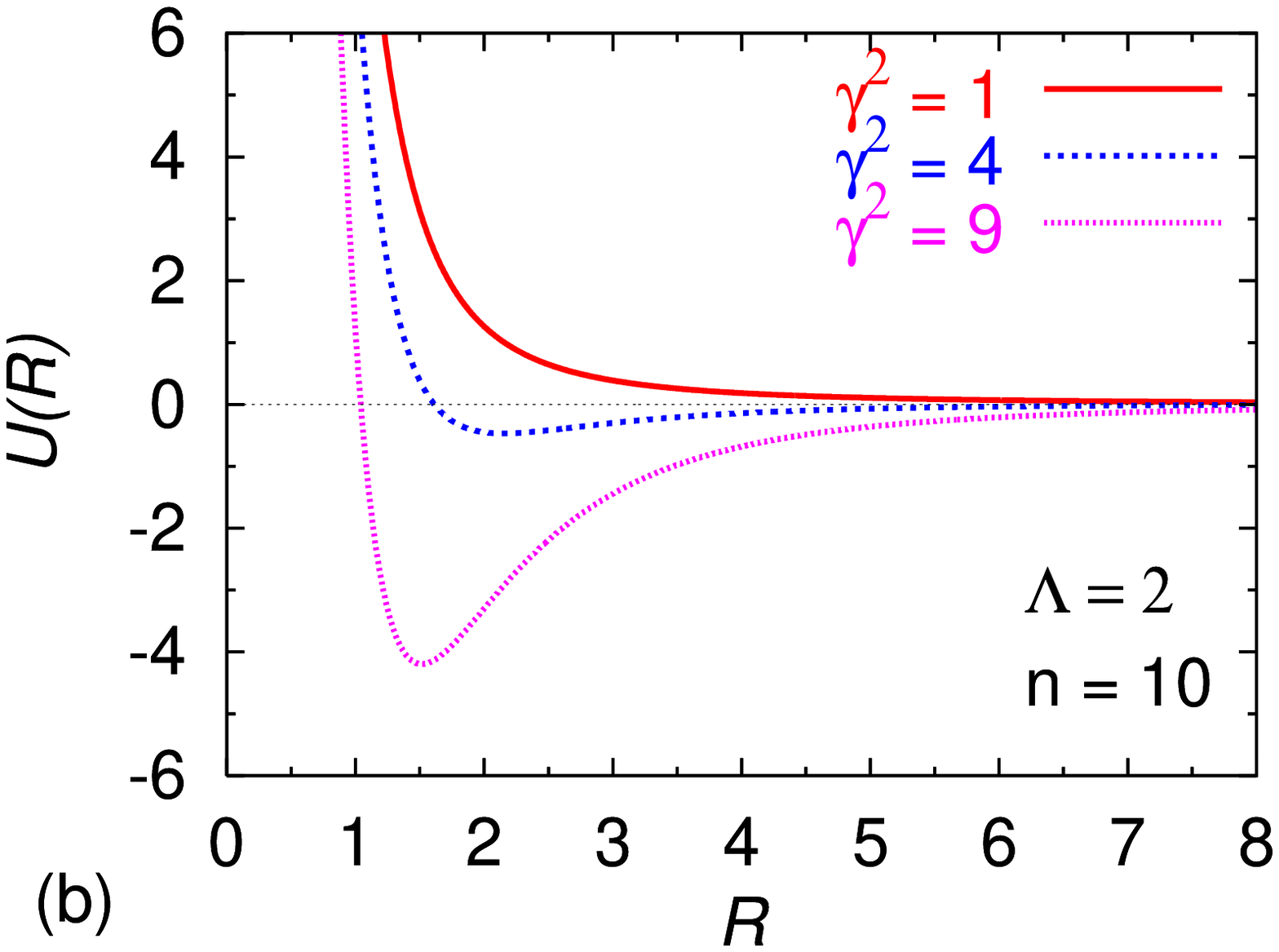}
\end{center}

\caption{The effective potential $U(R)$ of Eq.
 (\ref{ef1}) vs.
$R$  in
dimensionless  units for $\Lambda=2$, (a)  $\gamma=2$ and
$n=40,10,0,-0.4,$ and
-1
(upper to
lower curves) and for
(b) $n=10$ and $\gamma=1, 2,$ and $3$
(upper to
lower curves).
}

\end{figure}

In Figs. 1 (a) and (b) we plot $U(R)$ vs $R$ of  Eq. (\ref{ef1})
in dimensionless units
for different values of
$n$, $\Lambda$ and $\gamma$. We exhibit   $U(R)$
vs $R$ for different $n$ for $\Lambda= \gamma =2$ in Fig. 1 (a). The 
same
for different $\gamma$ and $\Lambda=2$ and $n=10$ are  exhibited in Fig. 
1
(b). We find from Fig. 1 (a) that for a fixed $\gamma$ and $\Lambda$,
$U(R)$
has a minimum  for $n_{\mbox{lim}}> n> n_{\mbox{crit}}$ corresponding to 
a
stable bound state, where  $n_{\mbox{lim}}$ is positive (repulsive
system) and $n_{\mbox{crit}}$ is negative (attractive system). In Fig. 1
(a) there is a minimum for $n=10, 0$ and $-0.4$ and no minimum for 
$n=40$
and $-1$. The reason for the nonexistence of bound state for $n>
n_{\mbox{lim}}$ and for
$n< n_{\mbox{crit}}$  are distinct.
The limiting condition   $n> n_{\mbox{lim}}$
corresponds to a highly repulsive system which cannot be bound in
the square-well potential with the given $\Lambda$ and $\gamma$. The
condition  $ n< n_{\mbox{crit}}$ corresponds to a highly attractive 
system
which collapses with the given  $\Lambda$ and $\gamma$.
In Fig. 1 (a) for
$n=40$ the system is unbound and for $n=-1$ it is unstable to collapse.
In Fig.  1 (a) we see
that as the attractive nonlinearity $|n|$ is increased,
one
of the walls of the well for small $r$  is gradually lowered and for a
sufficiently large
attraction
this wall is completely absent and the condensate collapses into the
infinitely deep well at the center for $n=-1$ in Fig. 1 (a).

In Fig. 1 (b) we illustrate the effect of increasing the strength of the
square-well potential for a fixed nonlinearity $n=10$. This is done by
varying the depth of the square-well potential $\gamma^2$ for a fixed
range $\Lambda$. For $\gamma = 2$ and 3,  $U(R)$ has a minimum and
the attraction of the square-well
potential is sufficient to bind the repulsive system with $n=10$. 
However,
for $\gamma =1$ there is no minimum in $U(R)$ and the square-well
potential is too weak to bind this
system.

\section{Numerical Result}

We solve NLS  equation (\ref{d1}) numerically  for the square-well
potential using the split-step time-iteration method employing the
Crank-Nicholson discretization scheme described recently \cite{11a,11b}.
We calculated the solutions with real-time propagation. However, we
checked  that imaginary-time propagation also leads to consistent
result.
To obtain the normalizable solution, the real time iteration is
started with the known solution of the linear problem with
scaled nonlinearity $n=0$. Then during  time iteration the nonlinearity
$n$ is switched on slowly
until its desired value is attained.  The change in the parameter 
should be such that it does not greatly alter the eigenvalue of the 
Hamiltonian (after time propagation). 
We also calculated the
nonnormalizable bound states in the continuum for a positive $n$. To
obtain this solution, the time iteration is started with a constant wave
function for a finite positive $n$ with $\gamma=0$. Then in the course 
of
time iteration the strength $\gamma$ of the square-well potential is
switched on slowly until its  desired value is attained. If
stabilization upon time iteration could be achieved for the chosen
parameters one already obtains the required nonlinear bound state in the
square-well potential. In previous studies we compared critically the
present numerical
scheme for the time-dependent NLS equation with other approaches
\cite{11a,11b} including
the time-independent schemes \cite{adhi} and assured that the present
approach
leads to results with high precision not only for the NLS equation with
one
space variable but also for  NLS equations in two and three space
variables \cite{am}. The agreement between the results obtained with
real and imaginary time propagation also assures the correctness of our
results.

\begin{figure}

\begin{center}
\includegraphics[width=.97\linewidth]{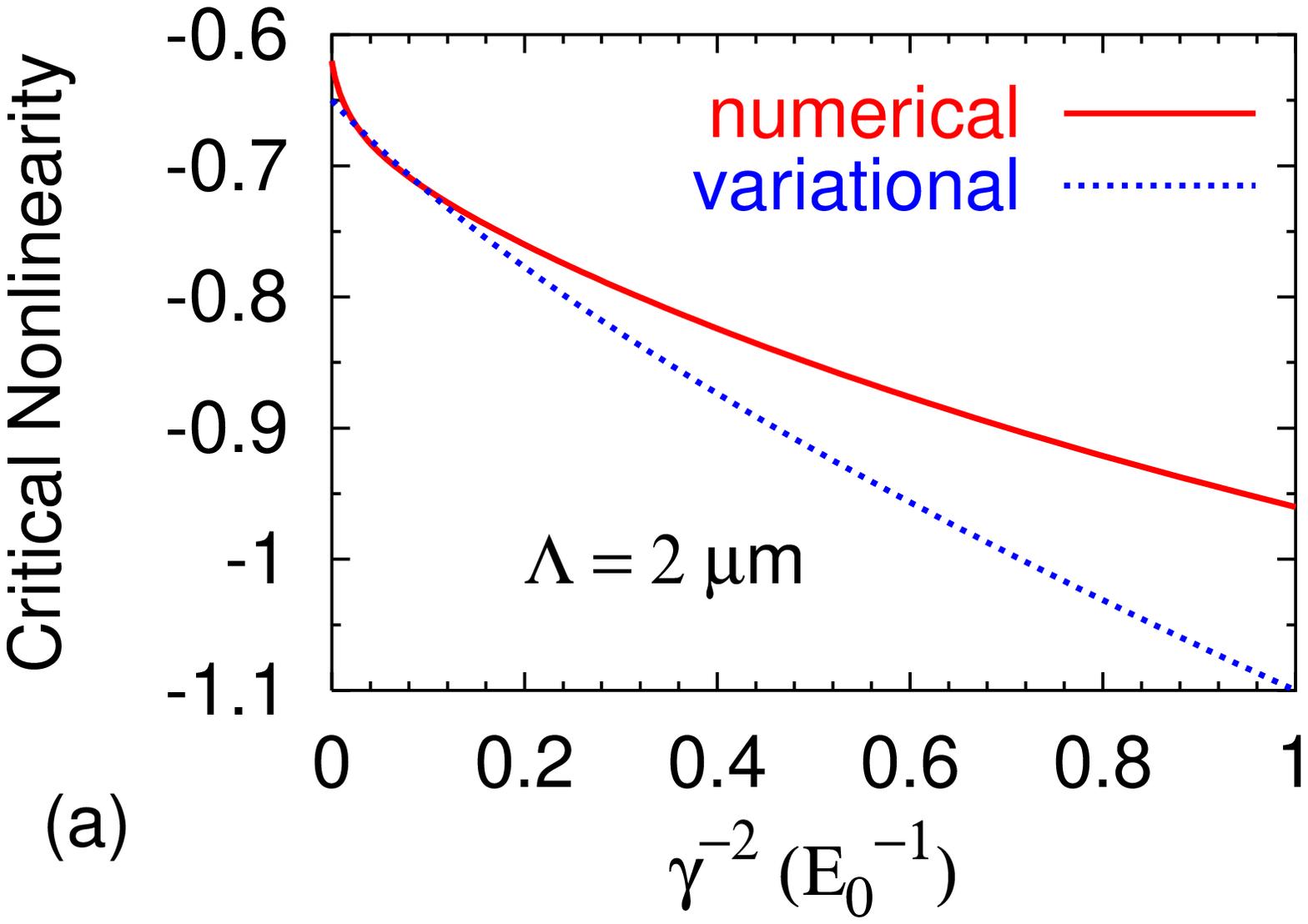}
\end{center}
\begin{center}
\includegraphics[width=.97\linewidth]{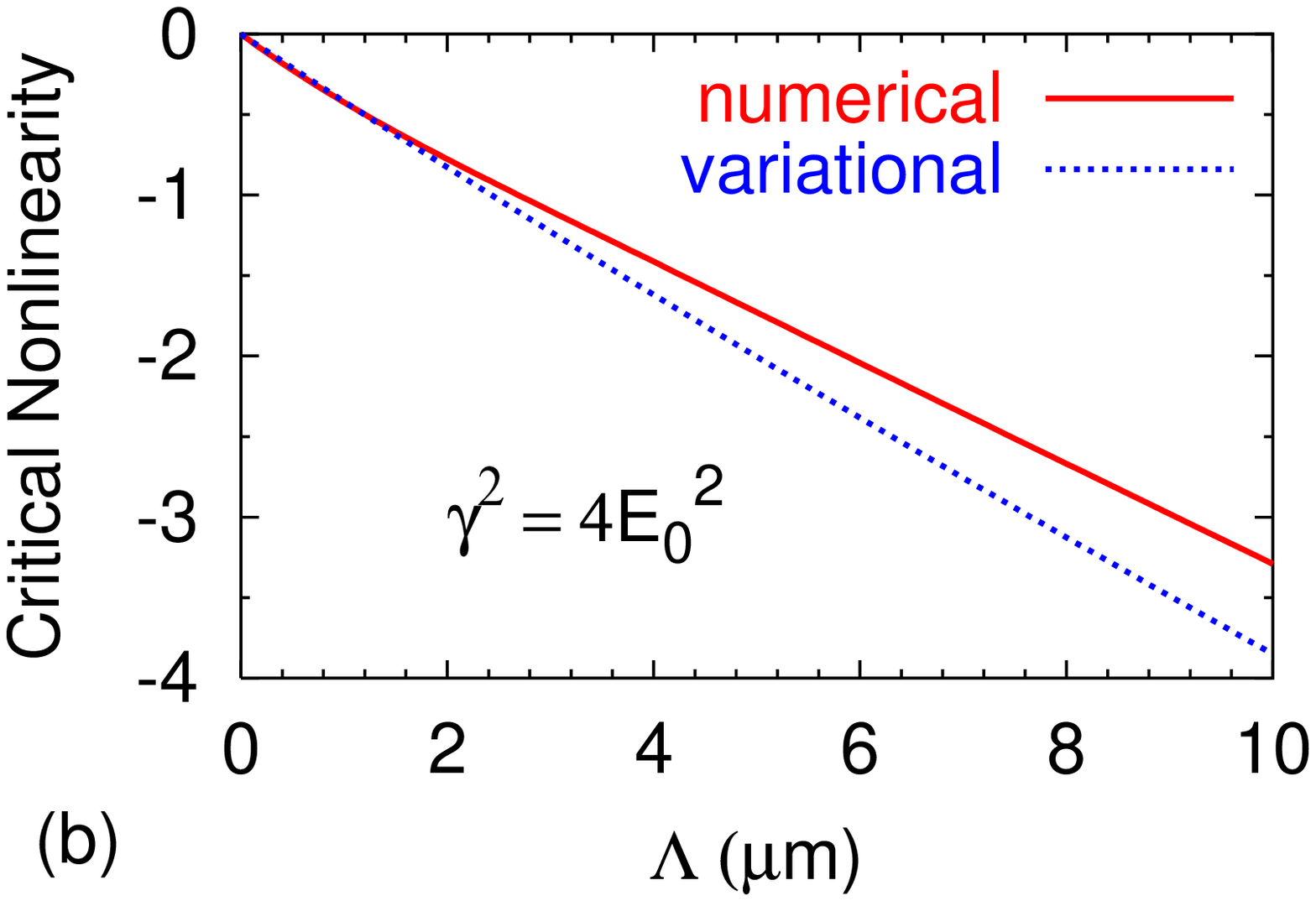}
\end{center}

\caption{Critical nonlinearity $n_{\mbox{crit}}$
(a) vs. $\gamma^{-2}$ for $\Lambda =2$ $\mu$m
and (b) vs. $\Lambda$ for $\gamma^2=4E_0\equiv 4\hbar\omega$: full line
-
numerical; dotted line - variational.
}

\end{figure}

\begin{figure}

\begin{center}
\includegraphics[width=1.\linewidth]{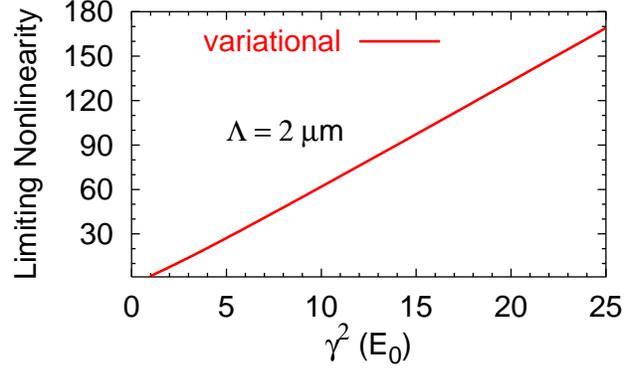}
\end{center}

\caption{Limiting  nonlinearity $n_{\mbox{lim}}$
 vs. $\gamma^{2}$ for $\Lambda =2$ $\mu$m.
}

\end{figure}

Although we calculate our results in dimensionless units,
typical parameters for an experimental realization can be easily 
obtained
therefrom for a particular atom. In the following we present results for
the Rb atom.
For  Rb let the length $l$
be 1 $\mu$m; for this  to be possible the reference frequency is
$\omega\approx 2\pi \times 38$ Hz. Consequently, the unit of energy is
$E_0\equiv \hbar
\omega \approx 1.57\times 10 ^{-13}$ eV.

\subsection{Normalizable States}

Stable normalizable bound
states are indeed found in all cases for various ranges of
parameters.
Some plausible properties of these
bound state are found in agreement with the above variational study. For
a given nonlinearity,
these bound states are
only formed  for a sufficiently strong square-well potential
determined by $\Lambda$ and $\gamma$. For weaker potentials, from the
wisdom obtained in variational calculation, the effective potential 
$U(R)$
does not have a minimum and there cannot be any bound state.
For a given square-well potential, bound states are found for
$n_{\mbox{lim}}
>n>n_{\mbox{crit}}$.

\begin{figure}

\begin{center}
\includegraphics[width=.8\linewidth]{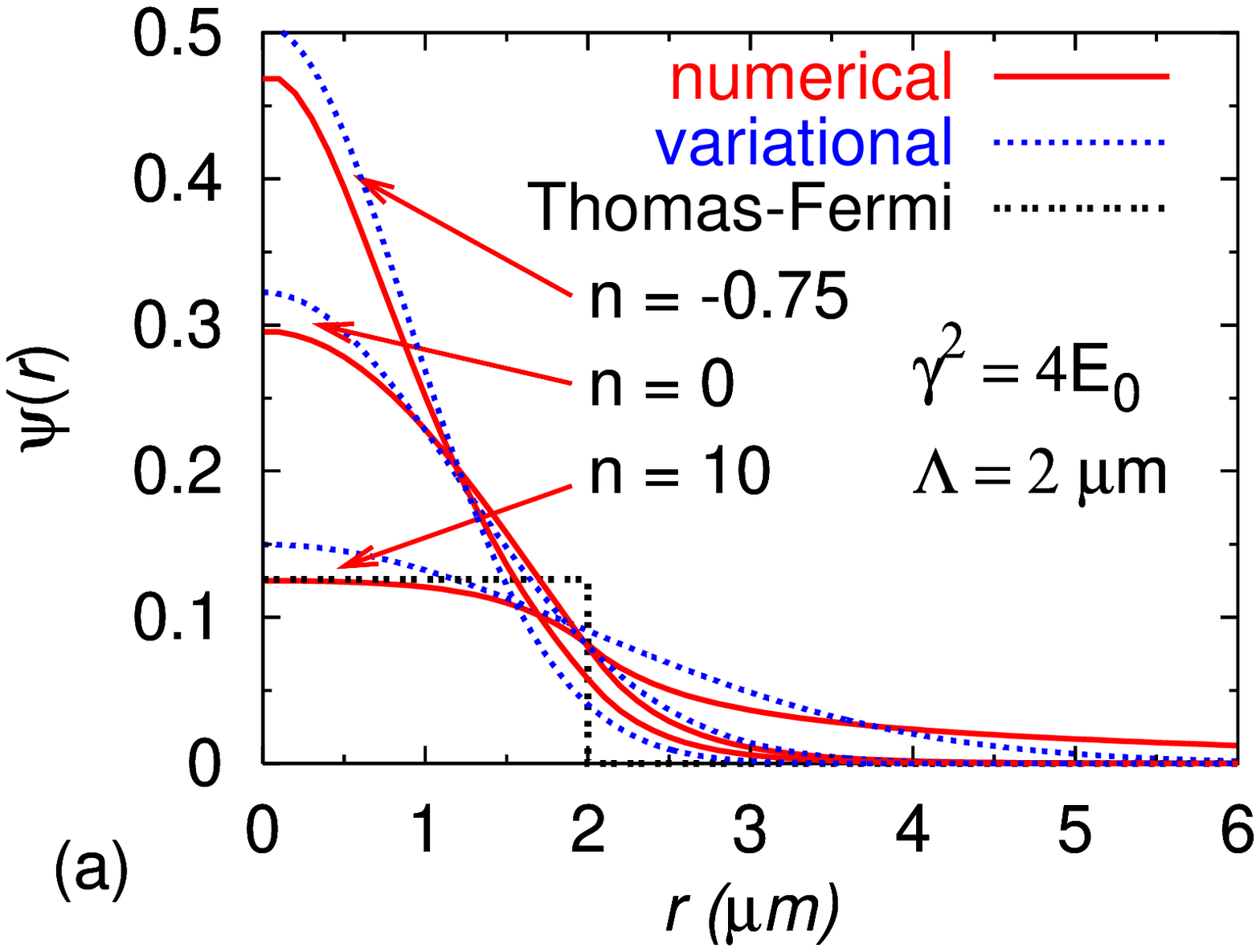}
\includegraphics[width=.8\linewidth]{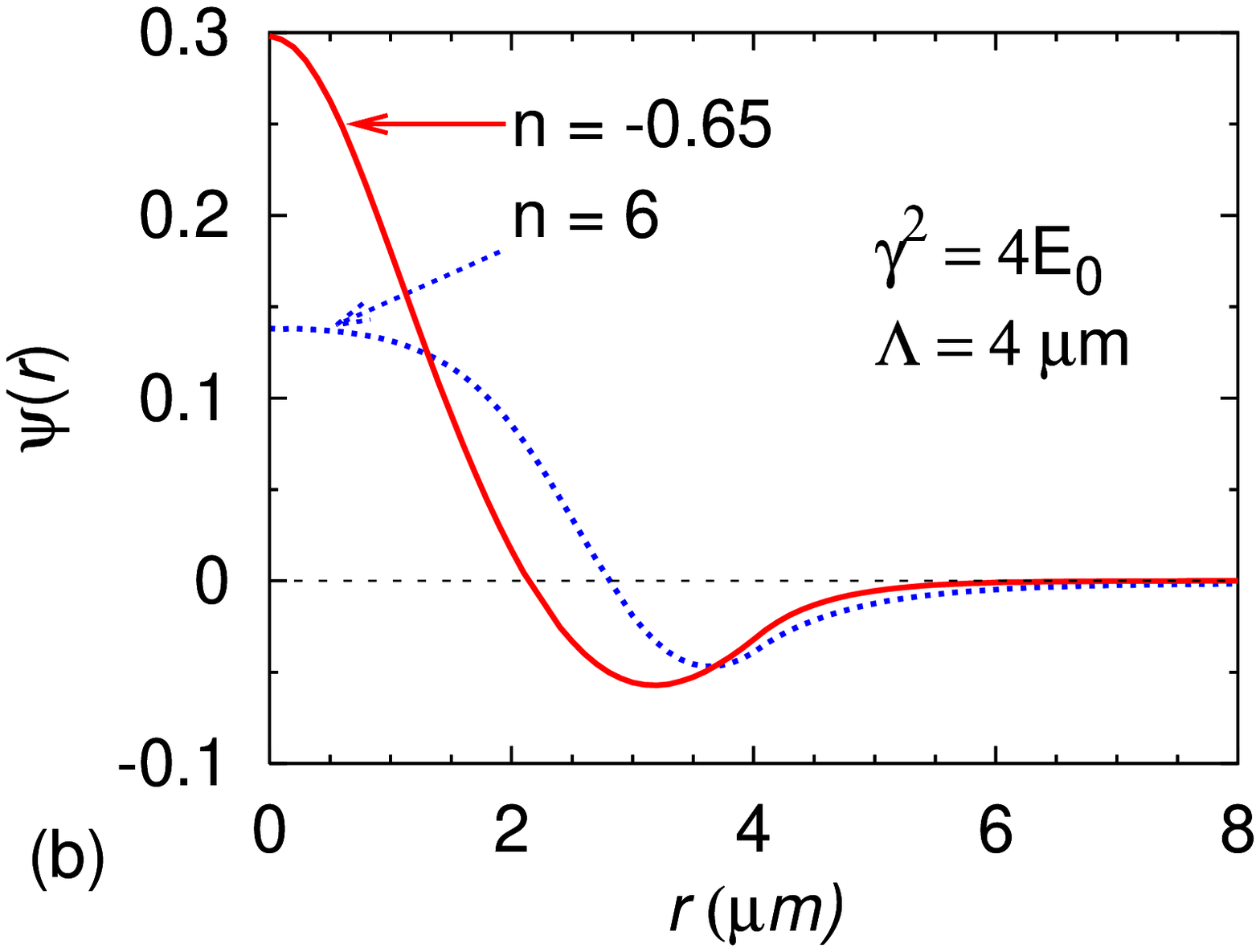}
\includegraphics[width=.8\linewidth]{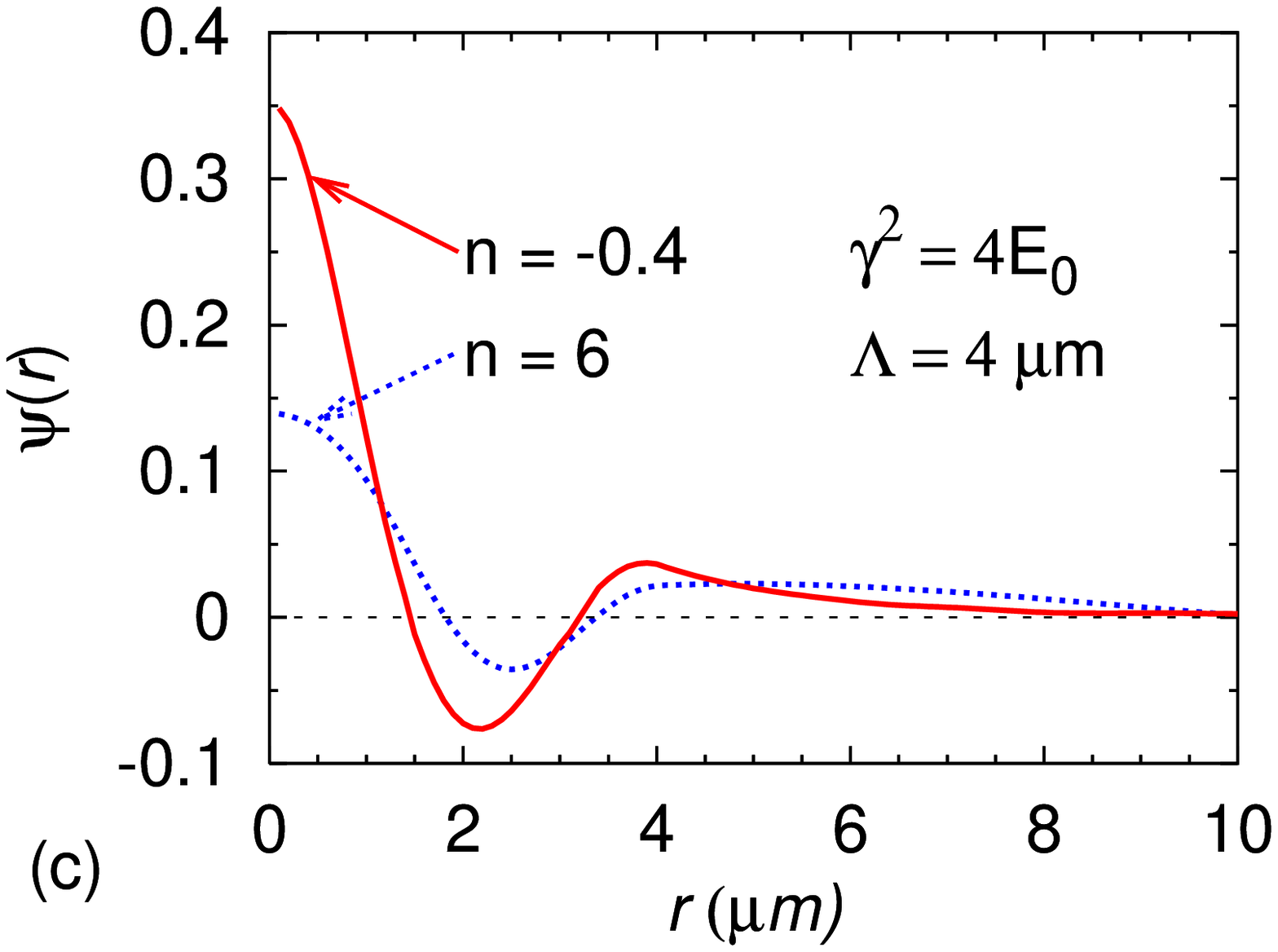}
\end{center}

\caption{
(a) Ground-state wave function  $\psi(r)$ of the
 NLS equation
(\ref{d2}) with the square-well potential
for  $\gamma^2 = 4E_0$, $\Lambda=2$ $\mu$m
and scaled  nonlinearity $n -0.75, 0$ and 10
(upper to lower curves). In all cases the numerical and variational 
results are
shown, in addition, for $n=10$ the TF approximation
is also
shown.
(b) Same for the first excited soliton (numerical calculation only)
with $\gamma^2=4E_0$, $\Lambda=4$ $\mu$m and
$n=-0.65$ and 6. (c)  Same for the second  excited soliton (numerical
calculation only)
with $\gamma^2=4E_0$,
$\Lambda=4$ $\mu$m and
$n=-0.4$ and 6.
}
\end{figure}

It is difficult to obtain
the limiting  nonlinearity  $n=n_{\mbox{lim}}$
numerically as this corresponds to a state with zero
energy
which extends to a very large $r$. However, the critical value
$n=n_{\mbox{crit}}$  can be obtained
numerically in a controlled fashion as the wave function is
highly localized in this limit.
In Figs. 2 we plot the critical nonlinearity for collapse
$n_{\mbox{crit}}$
for different parameters of the square-well potential. In Fig. 2 (a) we
plot $n_{\mbox{crit}}$  vs.  $\gamma^{-2}$ for $\Lambda=2$  $\mu$m 
whereas
in
Fig. 2 (b)  we
plot $n_{\mbox{crit}}$  vs.   $\Lambda$ for $\gamma^2 =4E_0$. In these
figures
we show results of variational and full
numerical calculations.  The variational calculation always leads to a
larger $|n_{\mbox{crit}}|.$ In case of the infinite harmonic potential
also, the variational estimate of  $|n_{\mbox{crit}}|$ is larger than 
the
result of full numerical calculation \cite{8}. For the infinite
square-well
potential with $\gamma^2=\infty $ and $\Lambda =2$ $\mu$m,
$n_{\mbox{crit}}=-0.62$; for the infinite parabolic potential
$n_{\mbox{crit}}=-0.575$ \cite{8,ad}. Because of the different shapes of
these two infinite potentials $n_{\mbox{crit}}$ is different in these 
two
cases.

\begin{figure}

\begin{center}
\includegraphics[width=.8\linewidth]{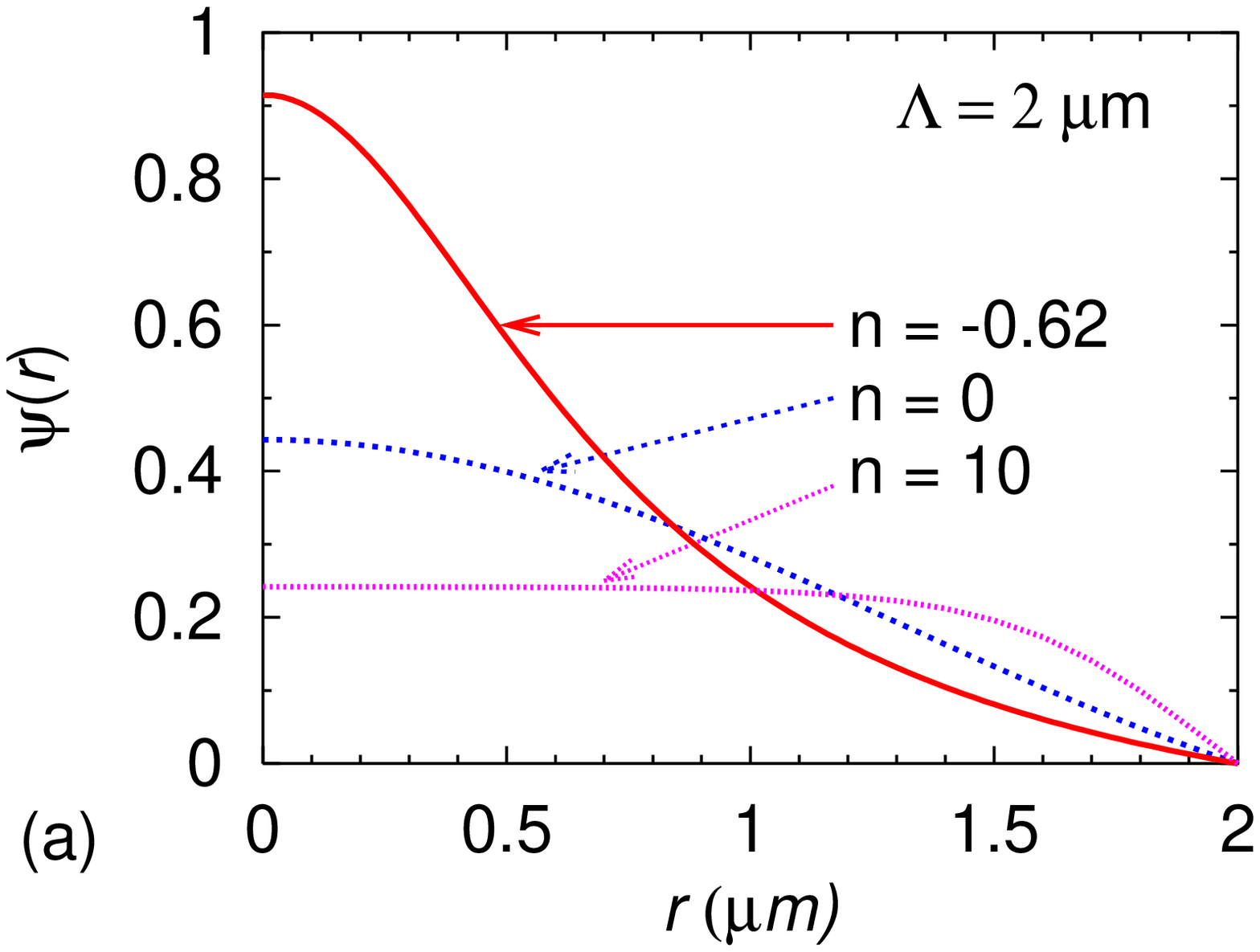}
\includegraphics[width=.8\linewidth]{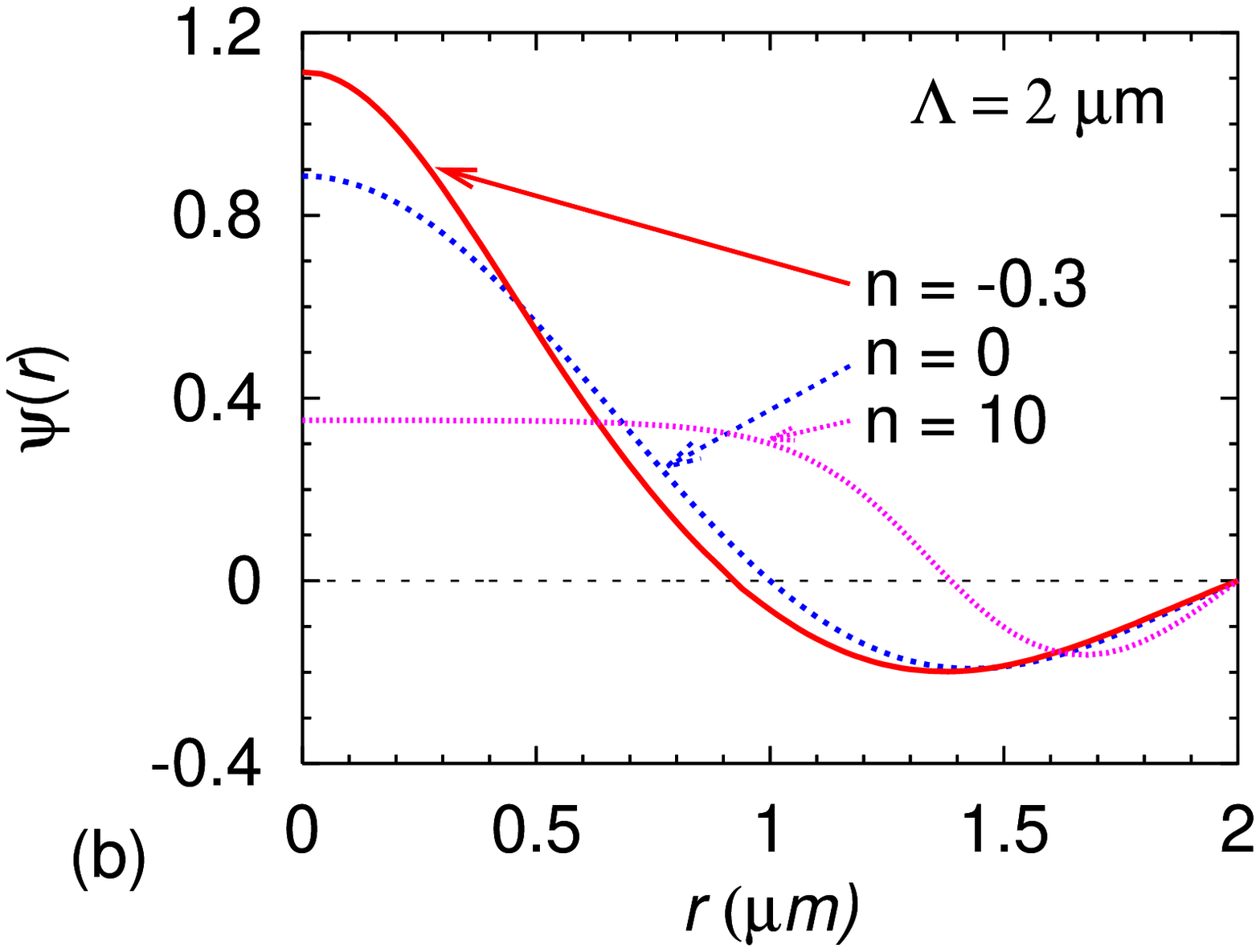}
\end{center}

\caption{
(a) Ground-state wave function  $\psi(r)$ of the
 NLS equation
(\ref{d2}) for  the infinite square-well potential of range
$\Lambda=2$ $\mu$m
and scaled  nonlinearity $n = -0.62, 0$ and 10
(upper to lower curves);
(b) same for the first excited soliton  for
$n=-0.3,0$ and 10.
}
\end{figure}

Although it is difficult to obtain $n_{\mbox{lim}}$ from a
numerical solution of the NLS equation, it is possible to obtain it
from the variational calculation. The limiting
nonlinearity
$n_{\mbox{lim}}$ corresponds to the
largest value of $n$ for which $U(R)$ has a minimum. In Fig. 3 we plot
$n_{\mbox{lim}}$ vs. $\gamma^2$ for $\Lambda =2$ $\mu$m. We see that
$n_{\mbox{lim}}$
increases linearly with the strength of the square well
$\gamma^2$. The variational calculation underbinds the system. The
repulsive nonlinearity destroys binding and a smaller repulsive
nonlinearity can destroy the weaker binding of the variational
model. Consequently, the variational limiting nonlinearity is smaller 
than
the actual limiting nonlinearity, which we verified in our numerical
calculation. The numerical calculation relies on the existence of a
localized wave function and it is difficult to calculate the limit when
this wave function extends to infinity and a localized wave function
ceases to exist. The
TF  wave function (\ref{tf1}) is always fully localized within $r <
\Lambda $  and is  inadequate for calculating the limiting
nonlinearity for $\mu \to 0$. The TF wave function  leads to a much
smaller value
$n_{\mbox{lim}}=
\Lambda^3 \gamma^2/6,$ based on imposing $\mu=0$ within the TF regime.

In Fig. 4 (a) we plot the wave function for the bound state of the NLS
equation
for a square-well potential with $\gamma^2=4E_0$, $\Lambda =2$ $\mu$m 
and
for
nonlinearities $n=-0.75, 0$ and 10. The system with $n=-0.75$ is
attractive, $n=0$ noninteracting, and $n=10$ repulsive.
In these cases results for both the full  numerical calculation and
variational approximation are shown. In addition, for $n=10$ the
TF
approximation (\ref{tf}) with $c=\mu=0$ is also shown.
In Fig. 4 (b) we show the wave
functions for the first  excited soliton  with a single node for 
$\gamma^2
=4E_0$
and $\Lambda =4$ $\mu$m and for $n=-0.65$ and 6.
In Fig. 4 (c) we show the wave
functions for the second  excited soliton
with two nodes for $\gamma^2 =4E_0$
and $\Lambda =4$ $\mu$m and for $n=-0.4$ and 6. In Figs. 4 we find that
the bound
state for the attractive system with negative $n$ values is more 
centrally
peaked than the bound state for the repulsive system with positive $n$
values. This is true for both ground and excited solitonic  states in 
the
finite
square-well potential
as well as for states in the infinite square-well
potential
studied below. The
central
peaking of the wave function for the attractive system corresponding to 
a
large central probability density is a consequence of the nonlinear
attraction.

Next we consider the infinite square-well potential of range $\Lambda$. 
In
this case the system remains confined in the region $0 \le r\le
\Lambda$. The
wave function is zero outside this region: $r> \Lambda$. We illustrate 
the
wave
functions in this case for different values of nonlinearity $n$ for the
ground state and the first excited soliton for $\Lambda =2$ $\mu$m. For
the
ground state
there is no node of the wave function for $0<r <\Lambda$. For the $j$th
excited soliton these are  $j$ nodes of the wave function in this 
region.
The wave functions for the ground state and the first excited soliton 
for
different  $n$ are shown in Figs. 5 (a) and (b), respectively.

\begin{figure}

\begin{center}
\includegraphics[width=.85\linewidth]{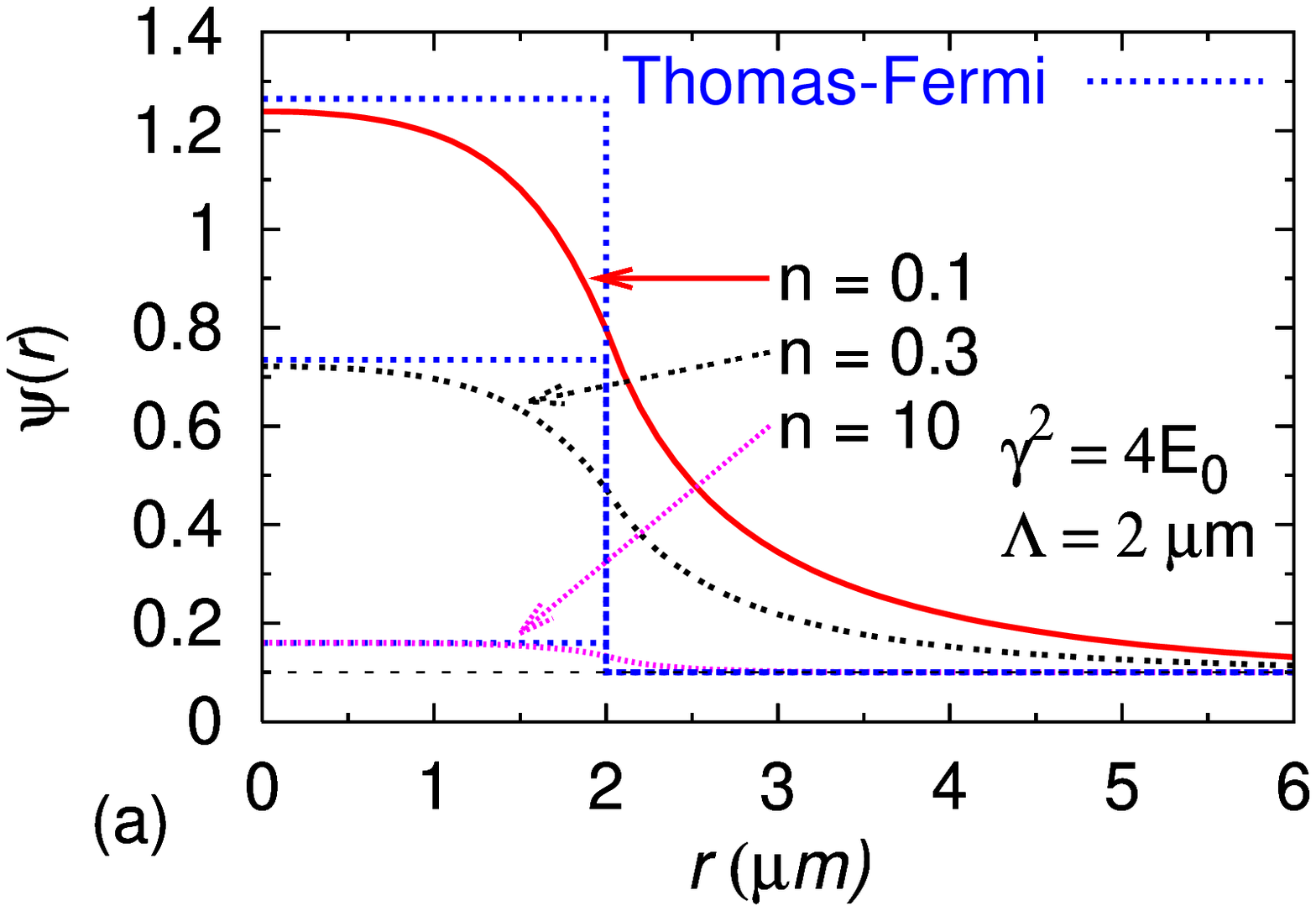}
\includegraphics[width=.85\linewidth]{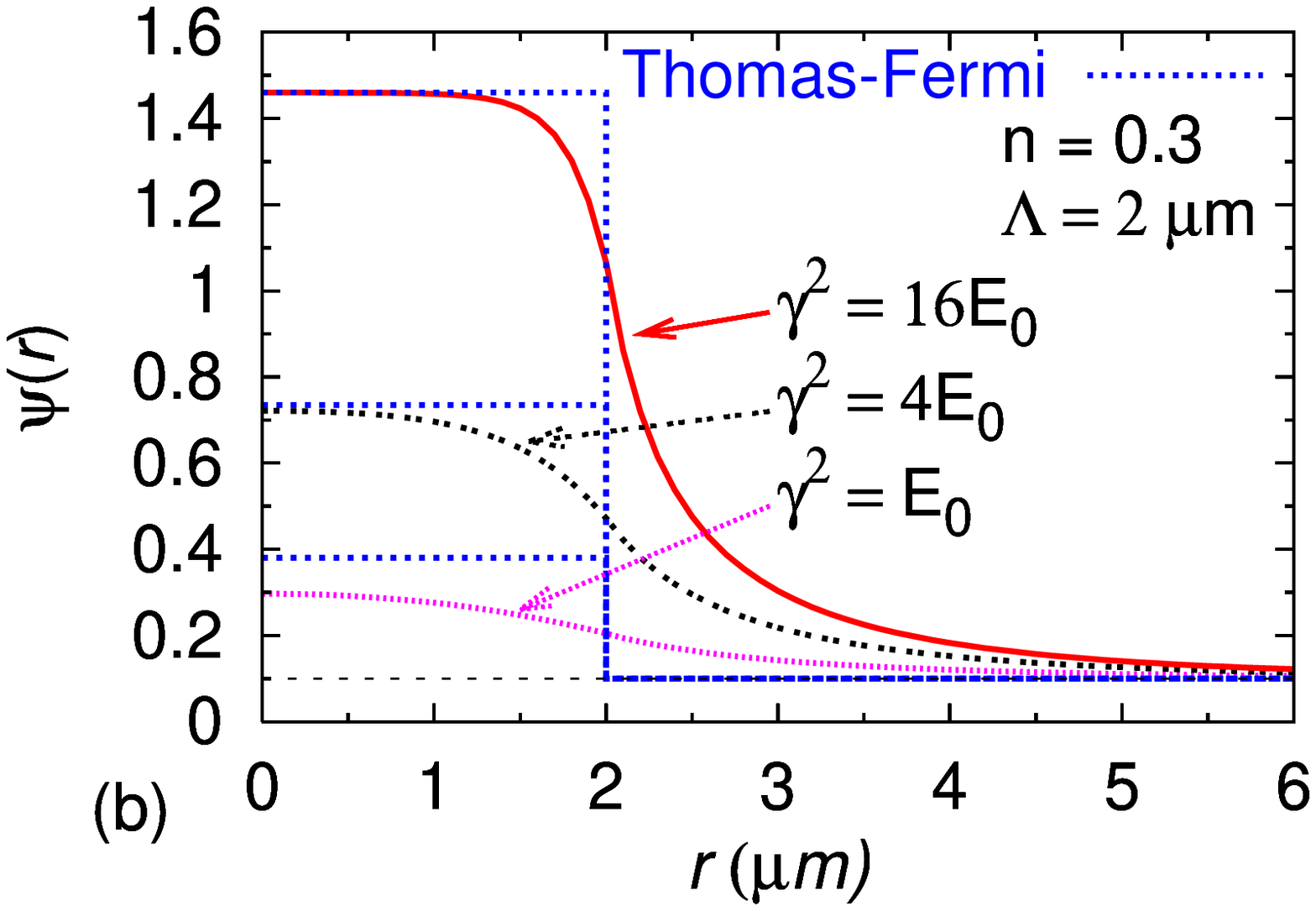}
\end{center}
\caption{Wave functions of the nonnormalizable states
$\psi(r)$
of the NLS equation
(\ref{d2}) for   (a) $\gamma^2=4E_0$, $\Lambda=2$ $\mu$m
and different $n$
values,
and for  (b) $n=0.3$, $\Lambda=2$ $\mu$m and different $\gamma$.
In both cases the
wave function tends to $0.1$ as $r\to \infty.$  In all cases the results
of the full numerical calculation and TF approximation
are shown.}
\end{figure}

\subsection{Nonnormalizable States}

Now we discuss the nonnormalizable solutions of Eq. (\ref{d1})
obtained for repulsive condensates.
No such states are found for attractive condensates with negative
$n$. To obtain  these solutions by time iteration of
Eq. (\ref{d1}) we start with the initial  constant solution 
$\psi(r)=0.1$
over all space for $V(r)=0$ and a fixed nonlinearity $n$. In the course 
of
time iteration the square-well potential is slowly introduced without
altering the nonlinearity until the full square-well potential is
obtained. The resulting wave function  of this calculation is plotted in
Figs. 6 (a) and
(b) together with the TF approximation. In Fig. 6 (a) we show
the wave function  for $\gamma^2 = 4E_0$, $\Lambda =2$  $\mu$m and 
$n=0.1,
0.3,$ and 10. In
Fig. 6 (b) we show     the same for $\Lambda=2$ $\mu$m, $n=0.3$ and
$\gamma^2 =E_0, 4E_0$ and 16$E_0$. From Fig. 6 (a) we find that for a
fixed $\gamma^2$
and $\Lambda$
the TF approximation becomes a better approximation  as the
nonlinearity increases; whereas from Fig. 6 (b) we find
that for a fixed $\Lambda$ and $n$  the TF approximation becomes a 
better
one as the strength of the
potential $\gamma^2$ increases.

The normalizable states discussed in the last subsection are true bound
states for negative $\mu$.  The nonnormalizable states occur at positive
$\mu$
and can be referred to as states in the continuum.  Similar
nonnormalizable states have been obtained in the study of the 1D NLS
equation with the square-well potential \cite{c2}.  However, there is 
the
possibility of physically observing the
nonnormalizable states with infinite norm as a transition from the
normalizable states with very large nonlinearity \cite{c2}.  An
increase of the scattering length $a$ via a Feshbach resonance 
\cite{fesh}
in a BEC may increase the nonlinearity $n (\equiv Na/l)$ indefinitely 
and
thus transform a normalizable state to a nonnormalizable one.  This
conclusion results from the following observation.  As the
normalization integral (\ref{n}) and the nonlinearity $g=8\pi n$ of Eq.
(\ref{d1}) are mixed in the present nonlinear model, an increase in $g$
can either be implemented my increasing the nonlinear term $g$ directly 
in
Eq.
(\ref{d1}) or by augmenting the normalization integral (\ref{n}) in the
same proportion.  Hence the nonnormalizable states with very large
(infinite) normalization could be a transition from normalizable states
with large nonlinearity.

\section{Discussion and Conclusion}

The study of the 1D NLS equation with the
simple square-well potential is of interest because of its simplicity 
and
intrinsic nonlinear nature. Its
interest in BEC and optics has  motivated recent
 investigation of
this problem \cite{c2}. In the 3D world, 1D
systems can only be achieved in some approximation and there are
nontrivial
differences between the solutions of the NLS in 1D and 3D
\cite{1}.

Hence we performed in this paper an investigation of the
3D
NLS equation with the square-well potential
using numerical and variational solutions. We find that the system 
allows
normalizable bound-state solutions for  $n_{\mbox{lim}}> n>
n_{\mbox{crit}}$, where the limiting nonlinearity $n_{\mbox{lim}}$
corresponds to a repulsive
(positive) limit beyond which the system cannot be bound
and the critical nonlinearity $n_{\mbox{crit}}$ to an attractive
(negative) nonlinearity below which the system collapses.
We calculated $n_{\mbox{crit}}$ for different parameters
of the square-well potential.
Many
results of this paper can be verified experimentally in BEC, where one 
can
make a square-well trap by joint magnetic and optical control
\cite{c2,step2}.
In
addition to the discrete  normalizable states we find
that
the 3D NLS equation with the square-well potential
also sustains nonnormalizable
states in the continuum of
interest in
nonlinear physics and mathematics. These states
cannot be obtained via  a
transition
from the  normalizable states.

\noindent{The work was supported in part by the CNPq and FAPESP
of Brazil.}


\end{document}